\documentclass[a4paper,11pt]{article}
\usepackage{graphicx,epsfig,pstricks,fancybox,fancyhdr,epic,rotating,color}
\usepackage{amsfonts}

\usepackage{graphicx,epsfig,pstricks,fancybox,fancyhdr,epic,rotating,color}
\usepackage{amsfonts}

\begin{document}

 \begin{center}

{\color{blue} \bf On the simplicity of numbers
}  
\vspace*{0.3cm}

{\color{blue}

{\bf Peter Minkowski}
\\
{\bf Albert Einstein Center for Fundamental Physics - 
ITP, University of Bern}
\\
\vspace{0.5cm}

{\color{magenta} \bf
Abstract} 
\vspace*{0.1cm} \\

{\color{blue}
\begin{tabular}{l}
The recent measurements of reactor $\ \overline{\nu}_{\ e} \ $
disappearance and its inter- \\
pretation in terms of the three light neutrino mixing angle
$\ \theta_{\ 13} \ $
by the \\
DAYA BAY $\ \rightarrow \ \theta_{\ 13} \ = 
\ \left ( \ 8.83^{\ + \ 0.81}_{\ - \ 0.88} \ \right )^{\ \circ} \ $
and \\
RENO \hspace*{0.76cm}
$\ \rightarrow \ \theta_{\ 13} \ =
\ \left ( \ 9.36^{\ + \ 0.88}_{\ - \ 0.96} \ \right )^{\ \circ} \ $
collaborations, gave rise to 
\\
this treatise, upon the hypothetical substitution
$\ \theta_{\ 13} \ \rightarrow \ \vartheta_{\ 9} \ = \ 9^{\ \circ} \ $.
\vspace*{0.1cm} \\
The latter angle ( $\ \vartheta_{\ 9} \ = \ 9^{\ \circ} \ $ )
is related to interesting algebraic
\\
properties of its periodic functions, which in turn have their origin
\\
in the discrete symmetry groups 
$\ S_{\ 5} \ = \ Z_{\ 2} \ \times \ A_{\ 5} \ $ and $\ A_{\ 5} \ $,
\\
the point groups associated with the regular d = 3 'Platonic bodies' :
\\
dodekahedron and ikosahedron. How these discrete groups may be
\\
related to dynamical symmetries of 
mass and mixing of light neutrino
\\
flavors is left open.
\end{tabular}
}
\vspace*{0.2cm}

{\color{green} 
15. April 2012
}
}
\end{center}

\newpage

\begin{center}

{\color{black} 
\section{Dodekahedron - Ikosahedron and pentagon related planar symmetry}
}
\end{center}

{\color{black}

\noindent
Algebraic identity for the angle $\ \vartheta_{\ 9} \ = \ 9^{\ \circ} \ $, 
related to the pentagon 

\vspace*{-0.1cm}
\begin{equation}
\label{eq:9-1}
\begin{array}{l}
\vartheta_{\ 9} \ = \ 9^{\ \circ} \ = \ 45^{\ \circ} \ / \ 5
\hspace*{0.2cm} : \hspace*{0.2cm}
s13 \ = \ \sin \ \left ( \ \vartheta_{\ 9} \ \right )
\hspace*{0.2cm} , \hspace*{0.2cm}
c13 \ = \ \sin \ \left ( \ \vartheta_{\ 9} \ \right )
\vspace*{0.2cm} \\ \hline \vspace*{-0.1cm} \\
s13 \ = 
\ \sqrt{\ \frac{1}{2} 
\ \left ( 
\begin{array}{l} 
\vspace*{-0.2cm} \\
\ 1 \ - 
\ \sqrt{\ 1 \ -
\ \left (
\ \begin{array}{c} 
\sqrt{5} \ - \ 1
\vspace*{0.2cm} \\ \hline \vspace*{-0.3cm} \\
4
\end{array}
\ \right )^{\ 2}
\ }
\end{array}
\ \right )
\ \hspace*{0.2cm}
}
\vspace*{0.3cm} \\
c13 \ =
\ \sqrt{\ \frac{1}{2} 
\ \left ( 
\begin{array}{l} 
\vspace*{-0.2cm} \\
\ 1 \ + 
\ \sqrt{\ 1 \ -
\ \left (
\ \begin{array}{c} 
\sqrt{5} \ - \ 1
\vspace*{0.2cm} \\ \hline \vspace*{-0.3cm} \\
4
\end{array}
\ \right )^{\ 2}
\ }
\end{array}
\ \right )
\ \hspace*{0.2cm}
}
\end{array}
\end{equation}

\noindent
From eq. \ref{eq:9-1} we can infer the half-angle nature of the
relation for \\
$\ s13 \ = \ \sin \ \left ( \ \vartheta_{\ 9} \ \right ) \ $

\vspace*{-0.1cm}
\begin{equation}
\label{eq:9-2}
\begin{array}{l}
\mbox{let :} \ \sin \ ( \ \varphi \ ) \ = 
\ \begin{array}{c}
\sqrt{5} \ - \ 1
\vspace*{0.2cm} \\ \hline \vspace*{-0.3cm} \\
4
\end{array}
\ =
\begin{array}{c}
1
\vspace*{0.2cm} \\ \hline \vspace*{-0.3cm} \\
1 \ + \ \sqrt{5}
\end{array}
\hspace*{0.3cm} \longrightarrow
\vspace*{0.1cm} \\
\varphi \ = \ \vartheta_{\ 18} \ = 
\ 2 \ \vartheta_{\ 9} \ = \ 18^{\ \circ}
\end{array}
\end{equation}

\noindent
To clarify the context of the following considerations let me state : 
in no way they are meant to 
imply exact , and less so 'deep' principles, which would
even in a clearly defined approximation, determine the exact values of the
angles, pertinant to the mixing matrix of the three light neutrino flavors

\vspace*{-0.1cm}
\begin{equation}
\label{eq:9-3}
\begin{array}{l}
\vartheta_{\ mn} \hspace*{0.2cm} ; \hspace*{0.2cm}
( \ m \neq n \ ) \ = \ 1,2,3 
\end{array}
\end{equation}

\noindent
or periodic functions thereof .

\noindent
This said, let me refer to the causes by recent events, that led to the
following treatise on 'The simplicity of numbers' :

\begin{description}

\item 1) First was the announcement by the DAYA-BAY collaboration on the
determination with remarkable statistical significane \\
of the (anti-)neutrino
mixing angle $\vartheta_{\ 13} \ $ in ref. \cite{DayaBay} .

\item 0) Really 1) was preceded by its announcement in correspondence by
Zhi-zhong Xing, also a member of the DAYA-BAY collaboration, who, 
after exchange of e-mails, also with Werner Rodejohann,
mentioned ref. \cite{Rodej} .

\item 2) Two papers deserve to be quoted in addition \cite{ZZXing} ,
\cite{SLuoZZXing} .

\end{description}

\newpage

\section{The simplicity of numbers}

From the Abstract of ref. \cite{DayaBay} , concerning the determination of 
$\ \vartheta_{\ 13} \ $ I quote : "A rate-only analysis finds

\vspace*{-0.1cm}
\begin{equation}
\label{eq:9-4}
\begin{array}{l}
\sin^{\ 2} \ \left ( \ 2 \ \theta_{\ 13} \ \right ) \ =
\ 0.092 \ \pm \ 0.016(stat.) \ \pm \ 0.005(syst.) 
\end{array}
\end{equation}

\noindent
in a three-neutrino framework." It is consistent with the more recent results 
of the RENO collaboration , ref. \cite{RENO}

\vspace*{-0.1cm}
\begin{equation}
\label{eq:9-4a}
\begin{array}{l}
\sin^{\ 2} \ \left ( \ 2 \ \theta_{\ 13} \ \right ) \ =
\ 0.103 \ \pm \ 0.013(sta.t) \ \pm \ 0.011(syst.) 
\end{array}
\end{equation}

\noindent
Using the notation, for experimental quantities 

\vspace*{-0.1cm}
\begin{equation}
\label{eq:9-5}
\begin{array}{l}
S13 \ = \ \sin \ \left ( \ \theta_{\ 13} \ \right )
\hspace*{0.2cm} ; \hspace*{0.2cm}
S213 \ = \ \sin \ \left ( \ 2 \ \theta_{\ 13} \ \right )
\vspace*{0.1cm} \\
X \ = \ \left ( \ S13 \ \right )^{\ 2} 
\hspace*{0.2cm} ; \hspace*{0.2cm}
Y \ = \ \left ( \ S213 \ \right )^{\ 2}
\end{array}
\end{equation}

\noindent
we have the functional relation

\vspace*{-0.1cm}
\begin{equation}
\label{eq:9-6}
\begin{array}{l}
X \ ( \ Y \ ) \ = \ \frac{1}{2} \ \left (
\ 1 \ - \ \sqrt{\ 1 \ - \ Y \ } \ \right )
\end{array}
\end{equation}

\noindent
Combining the errors in eq. \ref{eq:9-4} in quadrature we use

\vspace*{-0.1cm}
\begin{equation}
\label{eq:9-7}
\begin{array}{l}
\sin^{\ 2} \ \left ( \ 2 \ \theta_{\ 13} \ \right ) \ = \ 0.092 
\ \pm \ 0.017 \ \mbox{(comb)}
\end{array}
\end{equation}

\noindent
Thus we obtain 

\vspace*{-0.1cm}
\begin{equation}
\label{eq:9-8}
\begin{array}{l}
X \ ( \ 0.109 \ , \ 0.092 \ , \ 0.075 \ ) 
\ = \ \left ( \begin{array}{l}
0.0280360183234318  
\vspace*{0.1cm} \\ 
0.0235548300171362 
\vspace*{0.1cm} \\ 
0.0191153984582164 \end{array} \right ) \hspace*{0.2cm} \rightarrow
\vspace*{0.1cm} \\
\theta_{\ 13} \ = \ \left ( \begin{array}{l}
9.63898499442633^{\ \circ}
\vspace*{0.1cm} \\
8.8284100120672^{\hspace*{0.30cm} \circ}
\vspace*{0.1cm} \\
7.94708265426291^{\ \circ}
\end{array} \right )
\ \sim \ \left ( \ 8.828^{\ + 0.811}_{\ - 0.881} \ \right )^{\ \circ}
\end{array}
\end{equation}

\noindent
The deviation $\ \Delta9 \ = \ 9^{\ \circ} \ - \ \theta_{\ 12} \ $
is less than 0.22 times the 1 sigma error .

\noindent
Upon {\it choosing} $\ \vartheta_{\ 13} \ \equiv \ \vartheta_{\ 9}  
\ = \ 9^{\ \circ} \ $, 
the 'simplicity of numbers' lies in the algebraic expression for
$\ s13 \ \equiv \ \sin \ \left ( \ \vartheta_{\ 9} \ \right ) \ $
given in eq. \ref{eq:9-1} .

\newpage

{\color{blue}

\section{The experimental determination of the
neutrino mixing angles $\ \theta_{\ mn} \ ; \ m \ < \ n \ = \ 1,2,3 \ $
including theoretical analyses
}

\noindent
In their own right , as witnessed by refs. \cite{Rodej} - \cite{T2K}, the recent
weeks have seen a major step , experimentally, by the DAYA BAY results, 
reported in ref. \cite{DayaBay} , followed by the RENO collaboration in
ref. \cite{RENO} ,
of strong evidence for the influence of 
$\ \theta_{\ 13} \ $, given in eqs. \ref{eq:9-7} and \ref{eq:9-8} :

\vspace*{-0.1cm}
\begin{equation}
\label{eq:9-9}
\begin{array}{l@{\hspace*{0.0cm}}l@{\hspace*{0.1cm}}l}
X  =  \left ( \sin ( \ \theta_{\ 13} \ ) \right )^{\ 2} 
& = & 
0.0235548300171362 \ \left \lbrack \begin{array}{l} 
+  0.0044811883062956 
\vspace*{0.1cm} \\
-  0.0044394315589198 
\end{array} 
\right \rbrack
\vspace*{0.2cm} \\
{\color{black}
X_{\ 9} \ = \left ( \sin ( \ 9^{\ \circ} \ ) \right )^{\ 2}
}
& {\color{black} =} & 
{\color{black} 0.0244717418524232}
\end{array}
\end{equation}

\noindent
Going back in time from the communication by the DAYA BAY collaboration
\cite{DayaBay} we turn towards the accelerator long baseline results by the
T2K collaboration \cite{T2K} , which reports on limits inferred for
the quantity $\ Y \ = \ \left ( \ S213 \ \right )^{\ 2} \ =
\ \left ( \ \sin \ ( \ 2 \ \theta_{\ 13} \ ) \ \right )^{\ 2} \ $ 
defined in eqs. \ref{eq:9-5} and \ref{eq:9-7} comparing with the DAYA BAY
and RENO \cite{RENO} results

\vspace*{-0.4cm}
\begin{equation}
\label{eq:9-10}
\begin{array}{l@{\hspace*{0.1cm}}l@{\hspace*{0.1cm}}l@{\hspace*{0.0cm}}}
Y  =  \left ( \sin ( \ 2 \ \theta_{\ 13} \ ) \right )^{\ 2}
& = &
\begin{array}{l}
0.092 \ \pm \ 0.016(stat.) \ \pm \ 0.005(syst.) 
\hspace*{0.1cm} ; \hspace*{0.0cm} 
\begin{array}[t]{l@{\hspace*{0.0cm}}}
\mbox{DAYA} \\
\mbox{BAY}
\end{array}
\vspace*{0.1cm} \\
0.103  \ \pm \  0.013(stat.) \ \pm \  0.011(syst.)
\hspace*{0.1cm} ; \hspace*{0.1cm}
\mbox{RENO}
\end{array}
\vspace*{0.1cm} \\
{\color{black}
Y_{\ 18} \ = \left ( \sin ( \ 18^{\ \circ} \ ) \right )^{\ 2}
}
& {\color{black} =} & 
{\color{black} \left ( \ 3 \ - \ \sqrt{5} \ \right ) \ / \ 8
\ = \ 0.0954915028125263}
\end{array}
\end{equation}

\noindent
The 90 \% confidence limits for Y as reported by the T2K collaboration are
derived for $\ \delta_{\ CP} \ = \ 0 \ $ and separately for 
a) normal and b) inverted hierarchy of light neutrino masses

\vspace*{-0.1cm}
\begin{equation}
\label{eq:9-11}
\begin{array}{l}
\begin{array}{lllll}
\mbox{a)} & 0.03  <  \left ( \sin ( \ 2 \ \theta_{\ 13} \ ) \right )^{\ 2}
 <  0.28 
& \mbox{for normal hierarchy} & &
\vspace*{0.1cm} \\
& \delta_{\ CP} \ = \ 0 &  & ; & \mbox{T2K}
\vspace*{0.1cm} \\
\mbox{b)} & 0.04 < \left ( \sin ( \ 2 \ \theta_{\ 13} \ ) \right )^{\ 2}
 < 0.34
& \mbox{for inverted hierarchy} & &
\end{array}
\end{array}
\end{equation}

\noindent
We turn to the analysis of ref. \cite{FogliLisi} , which we first compare
with eq. \ref{eq:9-9} using the values corresponding to newly corrected reactor
fluxes at 1 $\sigma$ 

\vspace*{-0.1cm}
\begin{equation}
\label{eq:9-12}
\begin{array}{l@{\hspace*{0.0cm}}l@{\hspace*{0.1cm}}l}
X  =  \left ( \sin ( \ \theta_{\ 13} \ ) \right )^{\ 2} 
& = & 
0.0235548300171362 \ \left \lbrack \begin{array}{l} 
+  0.0044811883062956 
\vspace*{0.1cm} \\
-  0.0044394315589198 
\end{array} 
\right \rbrack
\vspace*{0.1cm} \\
& & \hspace*{4.5cm} \mbox{DAYA BAY}
\vspace*{0.1cm} \\
X  =  \left ( \sin ( \ \theta_{\ 13} \ ) \right )^{\ 2}
& = & 0.025 \ \pm \ 0.007 
\hspace*{0.2cm} ; \hspace*{0.2cm} \mbox{ref. \cite{FogliLisi}}
\vspace*{0.2cm} \\
{\color{black}
X_{\ 9} \ = \left ( \sin ( \ 9^{\ \circ} \ ) \right )^{\ 2}
}
& {\color{black} =} & 
{\color{black} 0.0244717418524232}
\end{array}
\end{equation}

} 

\newpage

{\color{blue}

\noindent
Next we consider the 3 $\sigma$ range of 
$\ Z  =  \left ( \sin ( \ \theta_{\ 12} \ ) \right)^{\ 2} \ $ in 
ref. \cite{FogliLisi}

\vspace*{-0.1cm}
\begin{equation}
\label{eq:9-13}
\begin{array}{l@{\hspace*{0.2cm}}l@{\hspace*{0.2cm}}l}
0.265 < Z  =  \left ( \sin ( \ \theta_{\ 12} \ ) \right )^{\ 2} 
< 0.364
& ; & \mbox{ref. \cite{FogliLisi}}
\vspace*{0.2cm} \\
{\color{black}
Z_{\ 36} \ = \left ( \sin ( \ 36^{\ \circ} \ ) \right )^{\ 2}
\ = \left ( \ 5 \ - \ \sqrt{5} \ \right ) \ / \ 8} & = & 
{\color{black}  0.345491502812526}
\end{array}
\end{equation}

\noindent
and similarly $\ ZZ  =  \left ( \sin ( \ \theta_{\ 23} \ ) \right)^{\ 2} \ $

\vspace*{-0.1cm}
\begin{equation}
\label{eq:9-14}
\begin{array}{l@{\hspace*{0.2cm}}l@{\hspace*{0.2cm}}l}
0.34 < ZZ  =  \left ( \sin ( \ \theta_{\ 23} \ ) \right )^{\ 2} 
< 0.64
& ; & \mbox{ref. \cite{FogliLisi}}
\vspace*{0.2cm} \\
{\color{black}
ZZ_{\ 45} \ = \left ( \sin ( \ 45^{\ \circ} \ ) \right )^{\ 2}
\ = \ 0.5}  &  & 
\end{array}
\end{equation}

\noindent
Let me stress once more, that there is no relation -- known to me -- 
between the number 5
and associated angles within the 'simplicity of numbers'

\vspace*{-0.1cm}
\begin{equation}
\label{eq:9-15}
\begin{array}{l}
\vartheta_{\ 13} \ \leftrightarrow \ 9^{\ \circ}
\hspace*{0.2cm} , \hspace*{0.2cm}
\vartheta_{\ 12} \ \leftrightarrow \ 36^{\ \circ}
\hspace*{0.2cm} , \hspace*{0.2cm}
\vartheta_{\ 23} \ \leftrightarrow \ 45^{\ \circ}
\end{array}
\end{equation}

\noindent
with the pentagon, dekagon as regular 2 dimensional structures, nor
with the dual pair dodekahedron - ikosahedron, as regular 3 dimensional ones, 
which could be associated
with a physically meaningful symmetry , limiting or absolute.

\noindent
Any such physical symmetry should also account for the
breaking pattern of the
unifying gauge group spin 10 or $\ D_{\ 5} \ $ , and then involves
roots and weigths , i.e. polyhedra in 5 dimensions 
( = the rank of $\ D_{\ 5} \ $ ) . In particular such symmtry relation
should include 'cristalline axes' thereof,
along which this group is broken as discussed in refs. 
\cite{venice3} , \cite{nuorigin08} .

\noindent
The decomposition of one family of the $\ D_{\ 5} \ $ 16 dimensional
spinor representation and its decomposition along the subgroup axes
corresponding to $\ D_{\ 5} \ \supset \ SU5 \ \times \ U1 \ $
can serve here as a guideline

\vspace*{-0.1cm}
\begin{equation}
\label{eq:9-16}
\begin{array}{l}
\left \lbrack \ 16 \ \right \rbrack \ = 
\ \left \lbrace \ 1 \ , \ 5 \ \right \rbrace
\ + \ \left \lbrace \ 10 \ , \ 1 \ \right \rbrace
\ + \ \left \lbrace \ \overline{5} \ , \ - 3 \ \right \rbrace
\end{array}
\end{equation}

\noindent
For completeness let me add two recent references on neutrino
flavor oscillations : ref. \cite{MaT7} on CP-phases and the discrete group
$\ T_{\ 7} \ $ and ref. \cite{Brancoreview} a review on  
leptonic CP violation .

}

\newpage

\begin{center}
\section*{A1 - Constructing the dodekahedron}
\end{center}

\vspace*{2.0cm}
\begin{center}
\hspace*{0.0cm}
\begin{figure}[htb]
\vskip -2.0cm
\hskip 2.0cm
\includegraphics[angle=-90,width=10.0cm]{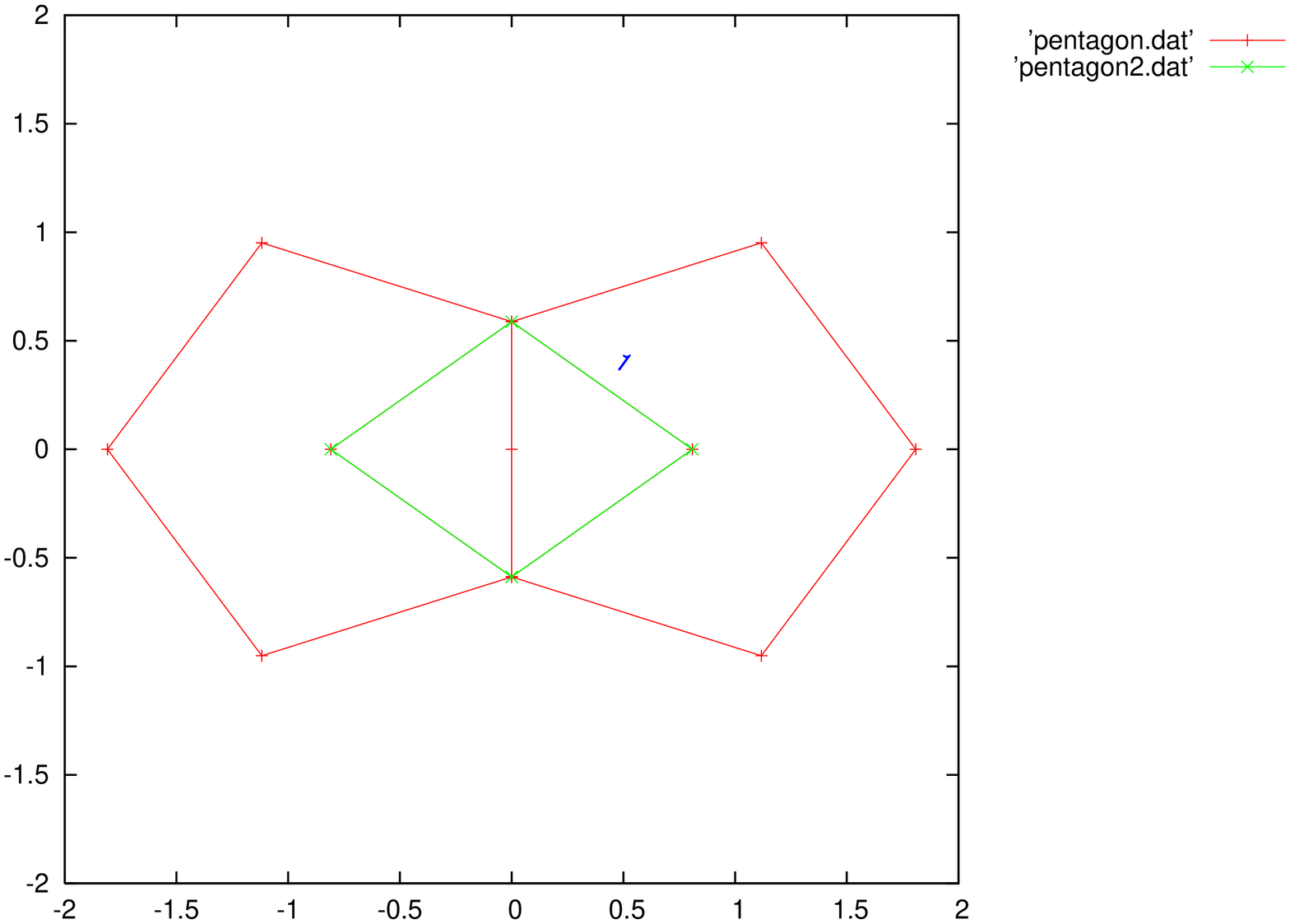}
\label{fig1}
\vspace*{+0.30cm}
{\small \bf \color{blue} \hspace*{2.5cm}
\begin{tabular}{l} Fig. 1 : 
Two pentagons in one plane
\vspace*{0.10cm} \\
\hspace*{1.4cm}
( x $\ \rightarrow $ ) , ( y $\ \uparrow $ ) , ( z $\ \odot $ )
\end{tabular}
}
\end{figure}
\end{center}

\noindent
We consider the x-axis to be parallel to the abscissa, the y-axis
parallel to the ordinate in Fig. 1 and the z-axis completing
a right oriented orthogonal basis, pointing out of the x-y plane of the
figure. The distance from the center of the pentagons to its
points is taken as unity, as shown in Fig. 1 .

\noindent
The two pentagon points lowest along the negative y-axis thus have the
cartesian coordinates ( eq. \ref{eq:9-2} )

\vspace*{-0.3cm}
\begin{equation}
\label{eq:9-17}
\begin{array}{l}
\mbox{PR(L)} \ : \ \left ( \ \pm \ \cos \ ( \ 18^{\ \circ} \ ) 
\  L \ , \ - 
\ \left ( \ \frac{1}{2} \ + \ \sin \ ( \ 18^{\ \circ} \ ) \ \right ) \ L
\ , \ 0 \ \right )
\vspace*{0.2cm} \\
\cos \ ( \ 18^{\ \circ} \ ) \ = 
\ \sqrt{
\ \begin{array}{c} 
5 \ + \ \sqrt{5}
\vspace*{0.2cm} \\ \hline \vspace*{-0.3cm} \\
8
\end{array}}
\hspace*{0.2cm} , \hspace*{0.2cm}
\sin \ ( \ 18^{\ \circ} \ ) \ = 
\ \begin{array}{c}
\sqrt{5} \ - \ 1
\vspace*{0.2cm} \\ \hline \vspace*{-0.3cm} \\
4
\end{array} 
\vspace*{0.1cm} \\
L \ = \ 2 \ \sin \ ( \ 2 \ \varphi \ ) \ = 
\ \sqrt{\ \begin{array}{c}
5 \ - \ \sqrt{5}
\vspace*{0.2cm} \\ \hline \vspace*{-0.3cm} \\
2
\end{array} \ }
\hspace*{0.2cm} ; \hspace*{0.2cm}
2 \ \varphi \ = \ 36^{\ \circ}
\end{array}
\end{equation}

\noindent
We rotate the right side pentagon around the y-axis by the angle $\ \psi \ $
in order to form an angle of $\ 108^{\ \circ} \ $ instead of
$\ 144^{\ \circ} \ $ between the nearest edge of the left side pentagon
and the rotated nearest edge of the right side one

\vspace*{-0.3cm}
\begin{equation}
\label{eq:9-18}
\begin{array}{l}
R_{\ y} \ ( \ \psi \ ) 
\ \left ( \begin{array}{l}
z
\vspace*{0.1cm} \\
x
\end{array} \right )
\ = 
\ \left ( \begin{array}{l}
\cos \ ( \psi ) \ z \ - \ \sin \ ( \psi ) \ x
\vspace*{0.1cm} \\
\sin \ ( \psi ) \ z \ + \ \cos \ ( \psi ) \ x
\end{array} \right )
\ =
\ \left ( \begin{array}{@{\hspace*{0.0cm}}r@{\hspace*{0.0cm}}}
- \sin(\psi) \ x
\vspace*{0.1cm} \\
\cos(\psi) \ x
\end{array} \right )
\end{array}
\end{equation}

\newpage

\noindent
The so rotated points $\ \mbox{PR}^{\ '} \ = \ R_{\ y} \ ( \ \psi \ ) 
\ \mbox{PL} \ $ and PL display the coordinates

\vspace*{-0.3cm}
\begin{equation}
\label{eq:9-19}
\begin{array}{l}
\left ( \begin{array}{l}
\mbox{PR}^{\ '} 
\vspace*{0.1cm} \\
\mbox{PL}^{\ }
\end{array} \right )
\ =
\ \left ( \begin{array}{lllll}
+ \cos (\psi) \ \cos (\varphi) 
& , & 
- \sin (\varphi) 
& , & 
- \sin (\psi) \ \cos (\varphi) 
\vspace*{0.1cm} \\
- \cos(\varphi) 
& , & 
- \sin(\varphi) 
& , & 
0
\end{array} \right )
\vspace*{0.1cm} \\
\varphi \ = \ 18^{\ \circ}
\end{array}
\end{equation}

\noindent
Thus we determine the sought angle between the two vectors 
displayed in eq. \ref{eq:9-19} from the relation

\vspace*{-0.3cm}
\begin{equation}
\label{eq:9-20}
\begin{array}{lll}
\cos \ ( \ \mbox{PR}^{\ '} \ , \ \mbox{PL} \ )
& = &
- \ \left ( \ \cos \ (\psi) \ \cos^{\ 2} \ (\varphi) 
\ - \ \sin^{\ 2} \ (\varphi) \ \right ) 
\vspace*{0.1cm} \\
& = &
- \ \sin \ (\varphi) \ =
\ \cos \ (108^{\ \circ})
\end{array}
\end{equation}

\noindent
which yields

\vspace*{-0.3cm}
\begin{equation}
\label{eq:9-21}
\begin{array}{lll}
\cos \ (\psi) \ = \ \tan^{\ 2} \ (\varphi) \ + 
\ \begin{array}{c}
\sin \ (\varphi) 
\vspace*{0.2cm} \\ \hline \vspace*{-0.3cm} \\
\cos^{\ 2} \ (\varphi)
\end{array}
& = & 
\begin{array}{c}
1 
\vspace*{0.2cm} \\ \hline \vspace*{-0.3cm} \\
\sqrt{5}
\end{array}
\vspace*{0.1cm} \\
\sin \ (\psi)
& = & 
\begin{array}{c}
2
\vspace*{0.2cm} \\ \hline \vspace*{-0.3cm} \\
\sqrt{5}
\end{array}
\end{array}
\end{equation}

\noindent
Eq. \ref{eq:9-21} establishes the algebraic simplicity of the
angle $\ \psi \ $ in the sense of 'simplicity of numbers' .

\noindent
$\ \psi \ $ itself is {\it not} a multiple of $\ 9^{\ \circ} \ $
however 

\vspace*{-0.3cm}
\begin{equation}
\label{eq:9-22}
\begin{array}{llr}
\psi \ = \ \mbox{arctg} \ (2) & = & 63.434948822922^{\ \circ} 
\vspace*{0.1cm} \\
\alpha \ = \ 180^{\ \circ} \ - \ \psi
& = & 116.565051177078^{\ \circ}
\end{array}
\end{equation}

\noindent
The angle between two planes of the dodekahedron,
which have one edge in common, called $\ \alpha \ $ or dihedral angle,
is defined as $\ \alpha \ = \ 180^{\ \circ} \ - \ \psi \ $ \\
in eq. \ref{eq:9-22} .

\begin{center}
\subsection*{Centering the faces}
\end{center}

\noindent
Having chosen the distance from the center of a pentagon to its points
of length unity as in Fig. 1, we turn
to the center points of the 12 pentagons forming the dodekahedron.

\noindent
Let CL be the center point of the unrotated left pentagon  ,
and $\ \mbox{CR}^{\ '} \ $ the center point of the
rotated right pentagon in Fig. 1. The associated two pentagons
shall be denoted $\ \mbox{penta} \ ( \ \mbox{L} \ ) \ $ and
$\ \mbox{penta} \ ( \ \mbox{R}^{\ '} \ ) \ $ respectively.

\noindent
They have coordinates, centered as in Fig. 1, using eq. \ref{eq:9-18} 

\vspace*{-0.4cm}
\begin{equation}
\label{eq:9-23}
\begin{array}{lll}
\mbox{CL} & : & 
\left (
\begin{array}{@{\hspace*{1.3cm}}llll@{\hspace*{3.7cm}}l@{\hspace*{0.4cm}}} 
- \ \cos \ ( \ 2 \ \varphi \ ) & , &
0 & , & 0 \end{array} \right )
\vspace*{0.1cm} \\
\mbox{CR}^{\ '} & : & 
 \left ( \begin{array}{lllll}
\cos \ ( \ \psi \ ) \ \cos \ ( \ 2 \ \varphi \ ) 
& , &
0 
& , & - \sin \ ( \ \psi \ ) \ \cos \ ( \ 2 \ \varphi \ )
\end{array}
\right )
\end{array}
\end{equation}

\noindent
The components of the outer normal to 
$\ \mbox{penta} \ ( \ \mbox{R}^{\ '} \ ) \ $, denoted
$\ \left ( \ \vec{n} \ \right )^{\ '} \ $, are

\vspace*{-0.3cm}
\begin{equation}
\label{eq:9-24}
\begin{array}{l}
\left ( \ \vec{n} \ \right )^{\ '} \ =
\ \left ( \ \sin \ ( \ \psi \ ) \ , \ 0 \ , \ \cos \ ( \ \psi \ )
\ \right )
\end{array}
\end{equation}

\newpage

\noindent
Thus we find the center of the dodekahedron intersecting the two straight lines
orthogonal to the faces , which lie in the $\ y \ = \ 0 \ $ plane.

\noindent
This gives rise to the equation by eq. \ref{eq:9-19}

\vspace*{-0.3cm}
\begin{equation}
\label{eq:9-25}
\begin{array}{l}
- \ \cos \ ( \ 2 \ \varphi \ ) \ =
\ - \ \lambda \ \sin \ ( \ \psi \ ) \ + 
\ \ \cos \ ( \ \psi \ ) \ \cos \ ( \ 2 \ \varphi \ )
\hspace*{0.2cm} \rightarrow
\vspace*{0.2cm} \\
\begin{array}{lll}
\lambda & = &  \cos \ ( \ 2 \ \varphi \ )
\ \begin{array}{c}
1 \ + \ \cos \ ( \ \psi \ ) 
\vspace*{0.2cm} \\ \hline \vspace*{-0.3cm} \\
\sin \ ( \ \psi \ )
\end{array}
\vspace*{0.2cm} \\
& = &
2 \ \left ( \begin{array}{c}
1 \ + \ \sqrt{5}
\vspace*{0.2cm} \\ \hline \vspace*{-0.3cm} \\
4
\end{array} \right )^{\ 2}
\ = \ 1.30901699437495
\end{array}
\end{array}
\end{equation}

\noindent
$ \lambda \ $ as defined in eq. \ref{eq:9-25} is the distance of any center
point of the (12) pentagons from the center of the dodekahedron, and thus is
the radius of the largest sphere inscribed within it, touching the faces
from inside at the twelve center points of pentagons.

\noindent
These 12 pentagon center points form 12 base points of an ikosahedron.
\vspace*{0.1cm} 

\noindent
We now translate the d=3 coordinates used for CL in eq. \ref{eq:9-23}
such that the center of the dodekahedron (to be) is at the origin.
Thus we introduce new coordinates 
$\ \vec{x}_{\ 1} \ = \vec{x} \ + \ \vec{T} \ $
according to the transformation

\vspace*{-0.3cm}
\begin{equation}
\label{eq:9-26}
\begin{array}{lll}
\vec{x}_{\ 1} & = & \left ( \hspace*{0.4cm}  x1 \hspace*{1.2cm} , 
\ y1 \ , \ z1 \ \right )
\hspace*{0.3cm} ; \hspace*{0.2cm}
\vec{x} \ = \ \left ( \ x \ , \ y \ , \ z \ \right )
\vspace*{0.2cm} \\
\vec{T} & = & 
\left ( \ \cos \ ( \ 2 \ \varphi \ ) \ , \ 0 \hspace*{0.4cm} , \hspace*{0.1cm} 
\lambda \hspace*{0.2cm} \right )
\vspace*{0.1cm} \\
 & = &
\left ( \ 0.809016994374947, \ 0 \ , \ 1.30901699437495 \ \right )
\end{array}
\end{equation}

\noindent
Next we go one step back to the 3d coordinates of $\ \vec{x} \ $
and determine the coordinates of the center point called $\ \mbox{CR}^{\ '} \ $
in eq. \ref{eq:9-23} , repeated below

\vspace*{-0.4cm}
\begin{equation}
\label{eq:9-27}
\begin{array}{lll}
\mbox{CL} & : & 
\left (
\begin{array}{@{\hspace*{1.3cm}}llll@{\hspace*{3.7cm}}l@{\hspace*{0.4cm}}} 
- \ \cos \ ( \ 2 \ \varphi \ ) & , &
0 & , & 0 \end{array} \right )
\vspace*{0.1cm} \\
\mbox{CR}^{\ '} & : & 
 \left ( \begin{array}{lllll}
\cos \ ( \ \psi \ ) \ \cos \ ( \ 2 \ \varphi \ ) 
& , &
0 
& , & - \sin \ ( \ \psi \ ) \ \cos \ ( \ 2 \ \varphi \ )
\end{array}
\right )
\end{array}
\end{equation}

\newpage

\vspace*{2.0cm}
\begin{center}
\hspace*{0.0cm}
\begin{figure}[htb]
\vskip -2.0cm
\hskip 2.0cm
\includegraphics[angle=-90,width=10.0cm]{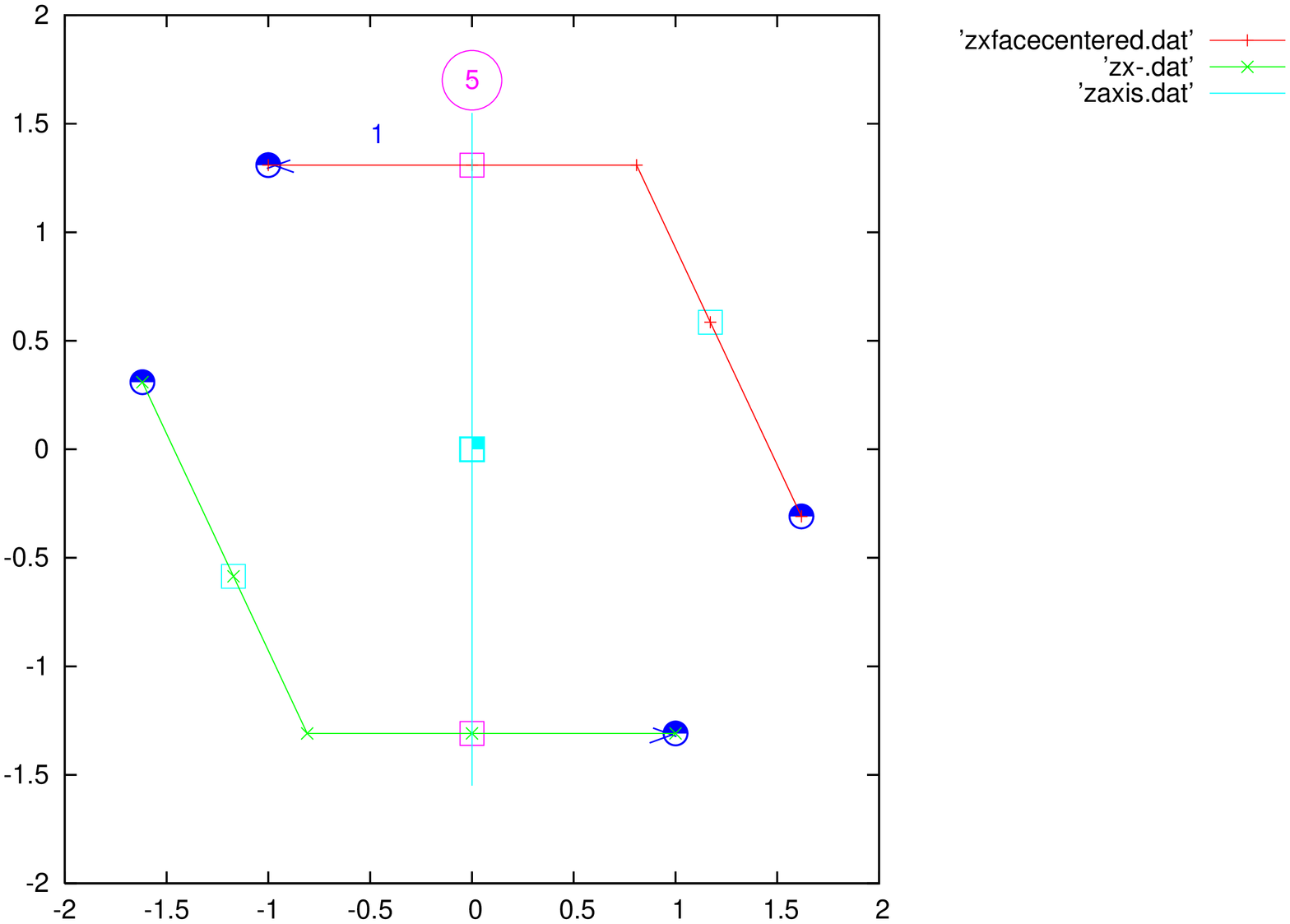}
\label{fig2}
\vspace*{+0.30cm}
{\small \bf \color{blue} \hspace*{2.5cm}
\begin{tabular}{l} Fig. 2 : 
The section $\ y \ = \ 0 \ $ as xz-plane
\vspace*{0.10cm} \\
\hspace*{1.4cm}
( x $\ \rightarrow $ ) , ( y $\ \otimes $ ) , ( z $\ \uparrow $ )
\end{tabular}
}
\end{figure}
\end{center}

\noindent
The coordinates x , y , z used in Fig. 2 represent the 
body centered system,  
pertaining to the translated vectors 
$\ \vec{x}_{\ 1} \ = \vec{x} \ + \ \vec{T} \ $, as defined in
eq. \ref{eq:9-26} . The suffix $\ _{1} \ $ is not explicitely indicated
for compactness of notation.

\noindent
Further in Fig. 2 the z-axis is drawn in cyan color . 
It is identical to the fivefold
axis and this is marked by the encircled number 5 .
The four half filled blue circles denote 4 points of the dodekahedron ,
from which the five rotations around the z-axis generate the 20 points or
corners of the dodekahedron.
\vspace*{0.1cm}

\noindent
The four empty quadrangles denote those center points of the 4 pentagons,
which lye in the $\ y \ = \ 0 \ $ plane drawn in Fig. 2 .
Upon performing the five symmetry rotations around the z-axis the two
quadrangle points, lying on it, remain invariant, whereas the other 
two generate the 10 points, which complete the twelve center points of the 
associated pentagons . These 12 points form basis corners of an ikosahedron .
\vspace*{0.1cm}

\noindent
Finally the quarter filled quadrangle in Fig. 2 marks the center of the
dodekahodron as well as of the associated ikosahedron.

\noindent Version 1
\hfill 31.03.2012

\noindent
Results of the RENO collaboration, ref. \cite{RENO}, added
\hfill 04.04.2012

\noindent Version 2 
\hfill 07.04.2012

\noindent Version 3 
\hfill 15.04.2012

}

\newpage

{\color{cyan} \large \bf

\thispagestyle{empty}

\vspace*{5.cm}

\begin{center}

Additional material

\end{center}

}

\newpage

\setcounter{page}{12}

\vspace*{2.0cm}
\begin{center}
\hspace*{0.0cm}
\begin{figure}[htb]
\hskip 2.0cm
\includegraphics[angle=0,width=10.0cm]{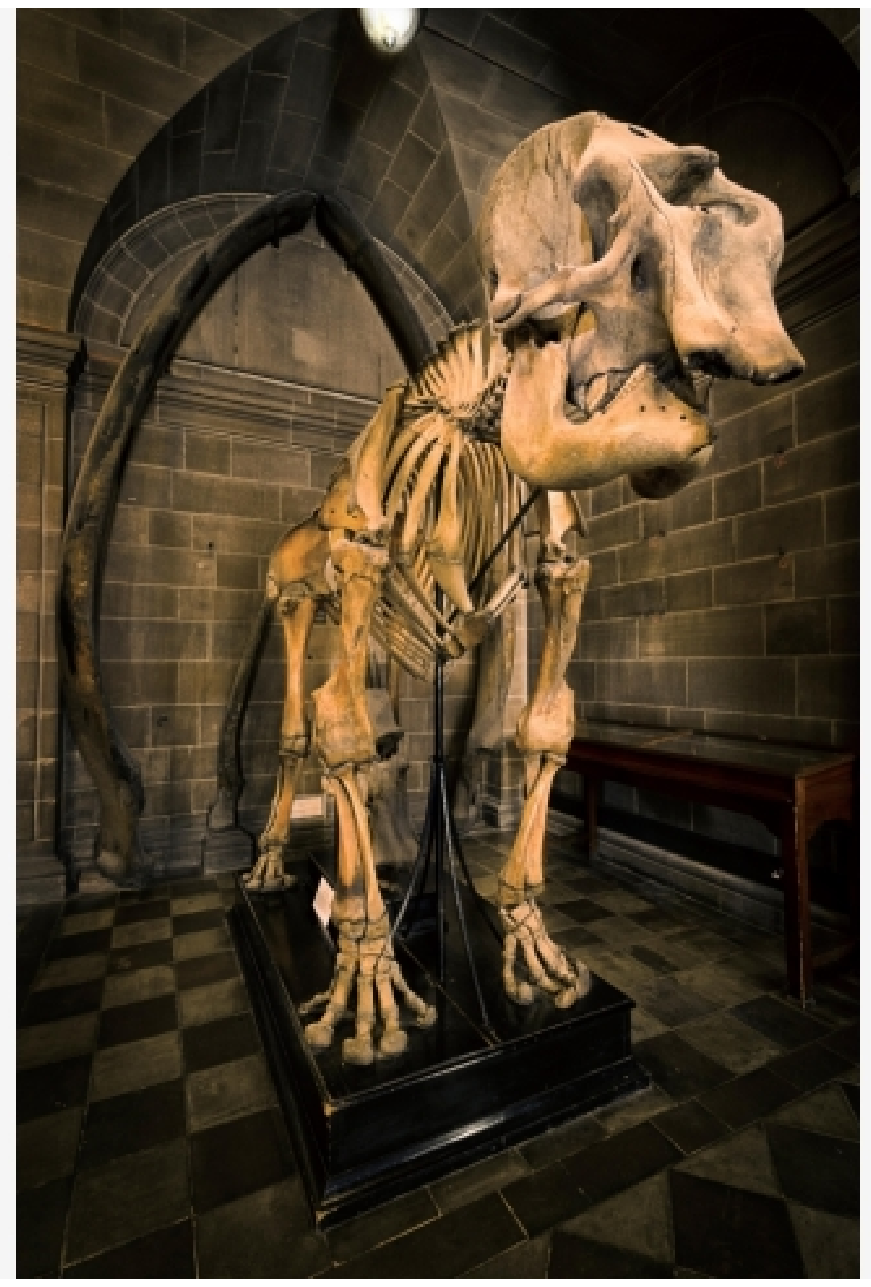}
\label{figb1}
\vspace*{+0.50cm}
{\tiny \bf \color{blue} \hspace*{2.7cm}
\begin{tabular}{l} Fig B1 : 
Skeleton of a an elephant or mammoth
\vspace*{0.10cm} \\
http://www.anatomy.mvm.ed.ac.uk/museum/$_{}$includes
\vspace*{0.10cm} \\
/img.php?crop=full/\&img=../upload/exhibits/elephant.jpg
\end{tabular}
}
\end{figure}
\end{center}


\newpage

\vspace*{-0.3cm}
\begin{thebibliography}{99}
\label{refmain}

\bibitem[1-2009]{Rodej} A. Adulpravitchai, A. Blum and
W. Rodejohann, 'Golden Ratio Prediction for Solar Neutrino Mixing',
March 2009. 14pp., New J.Phys.11 (2009) 063026, arXiv:0903.0531 [hep-ph] ,
and references cited therein.

\bibitem[2-2012]{DayaBay} F. P. An et al., DAYA-BAY Collaboration
'Observation of electron-antineutrino disappearance at Daya Bay', March 2012,
arXiv:1203.1669 [hep-ex] , and references cited therein.

\bibitem[3-2012]{ZZXing} Zhi-zhong Xing,
'Implications of the Daya Bay observation of
$\ \theta_{\ 13} \ $ on the leptonic flavor mixing structure and CP violation',
March 2012, 25pp., arXiv:1203.1672 [hep-ph] ,
and references cited therein.

\bibitem[4-2012]{SLuoZZXing} Shu Luo, Zhi-zhong Xing,
'Impacts of the observed $\ \theta_{\ 13} \ $ on the running
behaviors of Dirac and Majorana neutrino mixing angles and CP-violating
phases', March 2012, 15pp., arXiv:1203.3118 [hep-ph] ,
and references cited therein.

\bibitem[5-2011]{FogliLisi} G.L. Fogli, (Bari U. \& INFN, Bari) , E. Lisi,
(INFN, Bari) , A. Marrone, (Bari U. \& INFN, Bari) , 
A. Palazzo, (Munich, Tech. U., Universe) , A.M. Rotunno, (Bari U.), 
'Evidence of $\theta_{\ 13} \ > \ 0 $ from global neutrino data analysis',
June 2011, (Received Sep 1, 2011). 8pp., Phys.Rev.D84 (2011) 053007 ,
arXiv:1106.6028 [hep-ph] ,
and references cited therein .

\bibitem[6-2011]{T2K} T2K Collaboration: K.Abe et al.,
'Indication of Electron Neutrino Appearance from an Accelerator-produced
Off-axis Muon Neutrino Beam', Phys.Rev.Lett.107 (2011) 041801, 
arXiv:1106.2822v2 [hep-ex] .

\bibitem[6-2112]{grinst} B. Grinstein and M. Trott,
'An expansion for Neutrino Phenomenology',
arXiv:1203.4410 [hep=ph] .

\bibitem[7-2005]{venice3} P. Minkowski, 'Netrino oscillations  -- a historical
overview and its projection', contribution to "Neutrino Telescopes in Venice,
2005",
URL : http://www.mink.itp.unibe.ch in file venice3.pdf

\bibitem[8-2008]{nuorigin08} P. Minkowski, 'The origin of neutrino mass --
stations along the path of cognition',
Contribution to -- Discrete'08 --
Symposium on the Prospects in the Physics of Discrete Symmetries
11.-16. December 2008, IFIC, Valencia, Spain,
URL : http://www.mink.itp.unibe.ch in file nuorigin08.pdf

\newpage

\bibitem[9-2012]{RENO} Soo-Bong Kim for the RENO collaboration,
'Observation of Reactor Electron Antineutrino Disappearance in the RENO
    Experiment', arXiv:1204.062 [hep-ex] . 

\bibitem[10-2012]{MaT7} H. Ishimori, S. Khalil and E. Ma,
'CP Phases of Neutrino Mixing in a Supersymmetric B-L Gauge Model with 
$T_7$ Lepton Flavor Symmetry', arXiv:1204.2705 [hep-ph] .

\bibitem[11-2012]{Brancoreview}
G. C. Branco, R. Gonzalez Felipe and F. R. Joaquim,
'Leptonic CP violation', to appear in Reviews of Modern Physics ,
arXiv:1111.5332v2 [hep-ph] . 

\end{thebibliography}
\end{document}